\documentstyle[preprint,revtex]{aps}

\def\prl{Phys.\ Rev.\ Lett.\ }

\def\prb{Phys.\ Rev.\ B }
\begin{document}
\preprint{December 1992}
\begin{title}
Gapless spin-fluid ground state in a random, \\
quantum Heisenberg magnet
\end{title}
\author{Subir Sachdev and Jinwu Ye}
\begin{instit}
Departments of Physics and Applied Physics, P.O. Box 2157, \\
Yale University, New Haven, CT 06520
\end{instit}
\begin{abstract}
We examine the spin-$S$ quantum Heisenberg magnet with
Gaussian-random,
infinite-range exchange interactions. The quantum-disordered phase is
accessed by generalizing to $SU(M)$ symmetry and studying the large $M$
limit. For large $S$ the ground state is a spin-glass,
while quantum fluctuations produce a spin-fluid state for small $S$.
The spin-fluid phase is found to be
generically gapless - the average, zero temperature,
local dynamic spin-susceptibility obeys $\bar{\chi} (\omega ) \sim
\log(1/|\omega|) + i (\pi/2) \mbox{sgn} (\omega)$
at low frequencies.
\end{abstract}
\narrowtext

Random quantum spin systems offer a useful laboratory for studying the
fascinating interplay between strong interactions and
disorder. Though not as complex or intractable as metal-insulator
transition systems, they are still rich enough to display a host of
unusual physical phenomena. Moreover, they can be realized in a
number of experimental systems, many of which have been studied
intensively in recent years~\cite{birg,gabe,kagome,rosen,qcrit}.

It is useful to distinguish two different types of possible ground
states of a random quantum magnet: ({\em a\/}) a state with magnetic
long-range-order ($\langle \hat{\cal S}_i \rangle \neq 0$ where
$\hat{\cal S}_i$ is the spin operator on site $i$) which can be a
spin-glass, ferromagnet or an antiferromagnet; ({\em b\/}) a quantum
disordered (or `spin-fluid')
state in which $\langle \hat{\cal S}_i \rangle = 0$ due to
the presence of strong quantum fluctuations. Many properties of the
magnetically ordered phase can be described by a semiclassical
analysis. In contrast, the spin-fluid phase and its zero-temperature
phase transition
to the magnetically ordered phase are  intrinsincally quantum
mechanical, and their properties are only very poorly understood.
This paper shall mainly focus on the properties of the spin-fluid
phase.

In early studies of random-exchange spin-1/2
Heisenberg spin chains by a numerical
renormalization group method, Ma {\em et.al.\/} and others~\cite{chandan}
noted that the low
temperature spin susceptibility $\chi (T)$ behaved approximately
like $T^{-\alpha}$ with
$\alpha < 1$. This behavior, and their analysis, suggested that the quantum
disordered phase generically possessed gapless excitations: the low
energy excitations arose from a significant probability of finding
a pair of spins which were essentially decoupled from the rest of the
system, and with only a weak, mutual, effective exchange interaction.
Subsequently, the results of a numerical analysis by Bhatt and
Lee~\cite{bhatt_lee} of a
three-dimensional random-exchange spin-1/2
Heisenberg antiferromagnet could be well fit
with the same functional form with
$\alpha \approx 0.66$. Experiments~\cite{sip} on many
lightly doped semiconductors have also found similar behavior in the
low temperature spin susceptibility.
More recently, Doty and Fisher~\cite{doty,dsf} have obtained numerous exact
results on random quantum spin chains; in particular, Fisher~\cite{dsf}
proved that the random-exchange, spin-1/2 Heisenberg chain has $\chi \sim
1/(T \log^2 (1/T))$ and is gapless.

In this paper we introduce a new solvable, random-exchange, quantum
Heisenberg magnet - its solution reduces to the determination of the
properties of an integro-differential equation, which is a difficult,
though not impossible task. Our model possesses infinite-range exchange
interactions, and is thus a solvable limit which is complementary to
the spin chains. The spin-fluid phase of our
model is generically gapless; however the physical mechanism of the
gaplessness appears to be quite different from that of the random
spin chains. Which of these two limits is closer to realistic three
dimensional models remains an open question.
Finally, our model is expected to display a transition to a spin glass phase.
We have not yet succeeded in unraveling the nature of this transition
and that of the replica symmetry breaking in the spin-glass phase -
these are issues we hope to address in a future publication.

The main result discussed in this paper is that the
$T=0$, average, local dynamic spin susceptibility of our model
has the following
form over the entire quantum disordered phase:
\begin{equation}
\bar{\chi} ( \omega ) = X \left[ \log\left(\frac{1}{|\omega|}\right)
+ i \frac{\pi}{2} \mbox{sgn} ( \omega ) \right] + \cdots
\label{marginal}
\end{equation}
where $X$ is a constant to be determined below, and the omitted terms
are subdominant in the limit $|\omega | \rightarrow 0$.
A notable feature of this form is that it is identical to the
`marginal' Fermi liquid susceptibility proposed on phenomenological
grounds by Varma {\em et. al.\/}~\cite{marginal}
as a description of the electronic
properties of the cuprates.
It is not completely unreasonable to begin a study of
the low-lying spin fluctuations
in the cuprates by using the infinite-range quantum spin model described
below;
however, at present we have no arguments
which can determine whether, or how, the `marginal' spectrum will
survive in more realistic models with charge carriers
and finite-range interactions.
Nevertheless, to our knowledge,
ours is so far the only bulk model to
display the `marginal' spectrum over an entire phase,
and one might hope that
mathematical structure of the mean-field theory
is of broader significance.

We consider the ensemble of Hamiltonians
\begin{equation}
{\cal H} = \frac{1}{\sqrt{NM}} \sum_{i>j} J_{ij} \hat{\cal S}_i
\cdot \hat{\cal S}_j
\end{equation}
where the sum over $i,j$ extends over $N$ sites, the exchange
constants $J_{ij}$ are mutually uncorrelated and
selected with probability $P(J_{ij} ) \sim \exp ( - J_{ij}^2 / (2
J^2 ))$, the $\hat{\cal S}$ are the spin-operators of the group
$SU(M)$, and the states on each site belong to a representation
labeled by the integer $n_b$ ($n_b = 2S$ for $SU(2)$; more generally
$n_b$ is the number of columns in the the Young tableau of the
representation~\cite{np}).
This model has been considered previously by Bray and Moore~\cite{bm1}
for the group $SU(2)$: they found strong evidence in favor of the
presence of spin-glass order at $T=0$ for all values of
$S$. Accessing the spin-fluid phase
therefore requires considerations of groups
other than $SU(2)$:
following a technique which has been successful in clean
antiferromagnets\cite{np,self1},
we generalize to the group $SU(M)$ and study the phase
diagram in the $n_b - M$ plane. (We have also studied the properties
of
random $Sp(M)$~\cite{self2} magnets with results that are very similar to the
simpler $SU(M)$ case considered here). There are three interesting
limits in the $n_b - M$ plane:
\newline
({\em A\/}) $n_b \rightarrow \infty$, $M$ fixed: this is the
semiclassical limit and yields ground states well within the
magnetically ordered spin-glass phase.
\newline
({\em B\/}) $M \rightarrow \infty$, $n_b$ fixed: this takes us deep into the
spin-fluid phase.
\newline
({\em C\/}) $ M \rightarrow \infty$, $n_b / M = \kappa$ fixed: this
is in many ways the most interesting limit, because by varying $\kappa$
one can interpolate between the spin-glass and spin-fluid phases.
Moreover, one expects a phase-transition between these two ground
states at a critical value of $\kappa = \kappa_c$.

The structure of the $N\rightarrow \infty$ limit was discussed in
Ref.~\cite{bm1}. We express the partition function as a coherent-state
path-integral~\cite{np}, introduce $n$ replicas, average the partition
function, and the saddle-point reduces to the quantum mechanics of
$n$ replicas of a single spin;
assuming the saddle-point
is spin-rotation invariant (this is true in both the spin-fluid and
spin-glass phases) we obtain the single-site coherent-state path integral
$Z_0 = \int {\cal D} \hat{\cal S} \exp({\cal L})$ with
\begin{equation}
{\cal L} = S_B + \frac{J^2}{2M} \int_0^{1/T}
d \tau d \tau^{\prime} Q^{ab} ( \tau - \tau^{\prime} ) \hat{\cal S}^{a} ( \tau
)
\cdot \hat{\cal S}^{b} ( \tau^{\prime} )
\label{z0}
\end{equation}
and the self-consistency condition
\begin{equation}
Q^{ab} ( \tau - \tau^{\prime} )= \frac{1}{M^2} \langle
\hat{\cal S}^{a} (\tau )
\cdot \hat{\cal S}^{b} ( \tau^{\prime} ) \rangle_{Z_0}
\label{qab}
\end{equation}
Here $a,b = 1 \ldots n$ are replica indices, $\tau$, $\tau^{\prime}$ are
Matsubara times,
and $S_B$ is the single-spin
kinematic Berry
phase term~\cite{np}.
The Edwards-Anderson order parameter~\cite{hertz} for the spin-glass phase
is $q_{EA} = Q^{aa} ( \tau
\rightarrow \infty )$. Moreover,
$Q^{ab}$, $a \neq b$, is $\tau$-independent and
non-zero only in the spin-glass
phase~\cite{gold,ysr}.

An exact evaluation of $Z_0$ is clearly
not possible. We therefore consider the large $M$ limit, discussing first the
limit $(C)$ above. This is achieved by
the Schwinger boson realization of
$\hat{\cal S}$
\begin{equation}
\hat{\cal S}_{\mu}^{a\nu} = b_{\mu}^{a\dagger} b^{a\nu}~~~~~~;~~~~~~
\sum_{\mu} b_{\mu}^{a\dagger} b^{a\mu} = n_b
\label{bosonrep}
\end{equation}
where $b$ is a boson annhilation operator, $\mu,\nu = 1 \ldots M$.
In the large $M$ limit, Eqns~(\ref{z0},\ref{qab}) reduce to the following
equations for the boson Green's function $G_{B}^{ab} ( \tau ) = (1/M)
\sum_{\mu} \langle T ( b^{a\mu} ( \tau ) b_{\mu}^{b\dagger}(0) ) \rangle$
and its Fourier transform $G_{B}^{ab} ( i\omega_n )$
\begin{equation}
G_{B}( i\omega_n) = \left(-i \omega_n + \lambda - \Sigma_{B}
( i\omega_n ) \right)^{-1}
\label{dyson}
\end{equation}
\begin{equation}
\Sigma_{B}^{ab} ( \tau ) = J^2 G_{B}^{ab} ( \tau)
G_{B}^{ab} ( \tau) G_{B}^{ba} ( -\tau)
\label{boson}
\end{equation}
while $Q^{ab} ( \tau ) = G_{B}^{ab} ( \tau) G_{B}^{ba} ( - \tau)$.
Here $\lambda$ is a chemical potential set by the constraint $G^{aa}
( \tau = 0^{-} ) = \kappa$. These two equations can be combined into a
a single integro-differential equation for $G^{ab}_B ( \tau )$. We
also require that solutions satisfy conditions imposed by
the spectral representation of a boson Green's function: $G_B^{aa} (
z)$ is analytic for $\mbox{Im} ( z) > 0$, $\omega~ \mbox{Im} (G_B^{aa}
( \omega + i 0^{+} )) \geq 0 $ and $G_B^{aa} ( z ) \sim -1/z$ for
large $|z|$.
The replica-diagonal components of
Eqns (\ref{dyson},\ref{boson}) also bear a formal resemblance to a
perturbative solution
of the infinite-dimensional Hubbard model~\cite{antoine};
however there are some significant
differences which turn out to have dramatic
consequences in the nature of the solution.

We will focus here only on the spin-fluid phase, whence all correlations are
replica-diagonal, and replica indices will be dropped.
An immediate consequence of (\ref{dyson},\ref{boson}) is that the
zero-temperature
boson spectrum must be gapless! For suppose that the spectral weight
$\mbox{Im} (G_B ( \omega + i 0^{+}) ) = 0 $ for $|\omega | < \Delta$;
then
(\ref{boson}), expressed in real frequencies, implies
that $\mbox{Im} (\Sigma_B (\omega + i0^{+}) ) = 0 $
for $|\omega | < 3 \Delta$ -
this agrees with the real-frequency version of (\ref{dyson})
only if $\Delta = 0$.

Let us focus on the low-frequency behavior of $G_B$: assume that $G_B
( \omega )
\sim \omega^{\mu}$. Then from (\ref{boson}) we get $ \mbox{Im} (
\Sigma_B ( \omega )) \sim \omega^{2+3\mu}$. This can be consistent
with (\ref{dyson}) only if $\lambda = \Sigma_B ( \omega = 0)$ and
$\mu = -1/2$. As $G_B$ is analytic in the upper-half frequency plane, we write
\begin{equation}
G_B ( z) = \frac{i \Lambda e^{-i\theta} }{\sqrt{z}} +
\ldots~~~~~~~~~~ \mbox{Im}(z) > 0
\label{lowg}
\end{equation}
where $\Lambda > 0$. The positivity conditions on the spectral weight
require $0< \theta < \pi/2$. Inserting this into (\ref{boson}) we
find for $\mbox{Im} (z) > 0$ that
\begin{equation}
\Sigma_B (z) = \Sigma_B (0) + i \frac{J^2 \Lambda^3 \sin (2\theta)}{\pi}
e^{i \theta } \sqrt{z} + \cdots
\end{equation}
Finally, this is consistent with (\ref{dyson}) if $\lambda = \Sigma_B
( 0)$ and
\begin{equation}
\Lambda = \left( \frac{\pi}{J^2 \sin(2 \theta )} \right)^{1/4}
\end{equation}
The parameter $\theta$ remains undetermined. This is fortunate, as we
need a single degree of freedom to satisfy the boson-number
constraint $G_B ( \tau = 0^{-} ) = \kappa$. We will treat $\theta$ as
the independent parameter, with $\kappa ( \theta )$ a function to be
determined.
We expect $\kappa \rightarrow 0$, as $\theta \rightarrow 0$;
increasing $\theta$ therefore corresponds to increasing `spin'.
We can also determine the low-frequency behavior of the
spin-susceptibility $\chi ( \tau ) = Q^{aa} ( \tau )$; we find that
it has the form (\ref{marginal}) with the constant $X$ given by
\begin{equation}
X = \frac{( \pi \sin(2 \theta ))^{1/2}}{2J}
\end{equation}
We expect the low-frequency susceptibility to increase monotonically
with increasing `spin' $\kappa$, and therefore increasing $\theta$. However,
$X$ has a maximum at $\theta = \pi/4$. This leads us to conjecture
that the transition to the spin-glass phase occurs at $\theta = \pi
/4$ and only the range of values $0 < \theta < \pi/4$ correspond to
the spin-fluid phase. A second possibility, which cannot be ruled
out, is that there is a first-order transition to a spin-glass phase
at a value of $\theta < \pi/4$.

We have performed a detailed numerical study of Eqns
(\ref{dyson},\ref{boson}) to determine the complete frequency dependence of
Green's function. We chose a value of $\theta$, and
a trial form for $\mbox{Im}( G_B (
\omega + i 0^{+} ))$ whose low-frequency limit satisfies Eqn~(\ref{lowg}).
The real-frequency version of (\ref{boson})
expresses $\mbox{Im} ( \Sigma_B ( \omega + i 0^{+} ))$ as a double
convolution of $\mbox{Im}( G_B (
\omega + i 0^{+} ))$; these convolutions were performed by direct numerical
integration. The real part $\mbox{Re} ( \Sigma ( \omega + i 0^{+} ))$
was determined by a Kramers-Kronig transform, and $\lambda$ was set
at $\lambda = \Sigma_B (\omega = 0 )$. Finally $\mbox{Im}( G_B (
\omega + i 0^{+} ))$ was determined from (\ref{dyson}) and the whole
procedure was iterated, until the solution converged. The
singularities in $G_B$ and $\Sigma_B$ at low frequencies were
accounted for by performing the numerical integration in a variable
$x \sim \sqrt{\omega}$ at the integration end-points - this absorbed the
leading singularity. Subleading singularities were treated by using a
dual mesh-size in the integration - a very fine mesh ($x$ spacing $=
0.0003 \sqrt{J}$) was used at the
end-points and a coarse mesh elsewhere. Upto 1700 points were used in
the numerical integration. There was little difficulty in converging
to a solution for values of $\theta$ less than approximately
$\pi/6$; we are reasonably certain
that there are physically sensible solutions of
(\ref{dyson},\ref{boson}) for this range values of $\theta$. One such
solution, at $\theta = \pi/12$ is shown in Fig~\ref{fbos}
which was found to have $\kappa = 0.051$.
The numerical iteration became increasingly unstable with increasing
$\theta$ and did not converge to any smooth solution for large $\theta$.
Our numerical experience is consistent with the
conjecture that there are no physically sensible solutions for
$\theta > \pi/4$ - this is the range of values of $\theta$ where we
expect a spin glass phase.

A complementary picture of the spin-fluid phase can be obtained by
studying the large $M$ limit $(B)$. This takes $M\rightarrow \infty$
at fixed spin $n_b$ - one is then in a region of the phase diagram well
away from the transition to the spin-glass phase.
For technical reasons it is
also necessary to introduce of order $M$ {\em rows} in the Young
tableau of the spin representation; this is
discussed in some detail in Ref.~\cite{np}. We will focus on the
particle-hole symmetric representations which have $n_b$ columns and
$M/2$ rows as realized by the following operator decomposition
\begin{equation}
\hat{\cal S}_{\mu}^{a\nu} = \sum_{\alpha}
f_{\alpha\mu}^{a\dagger} f^{a\alpha\nu}
{}~~;~~\sum_{\mu} f_{\alpha\mu}^{a\dagger}
f^{a\beta\mu}  = \delta^{\beta}_{\alpha} M/2
\label{fermionrep}
\end{equation}
where $f$ is a fermion annhilation operator. The fermions carry
replica, spin, and `color' indices $\alpha,\beta = 1 \ldots n_b$.
The subsequent analysis parallels closely that for the bosons. The
fermion Green's function is replica and color diagonal and its only
non-zero component is
$G_{F} ( \tau ) = (1/M)
\sum_{\mu} \langle T ( f^{a\alpha\mu} ( \tau )
f_{\alpha\mu}^{a\dagger}(0) ) \rangle$. The only changes are that
Eqn~(\ref{boson}) is replaced by
\begin{equation}
\Sigma_F ( \tau ) = - J^2 n_b G_F^2 ( \tau ) G_F ( - \tau )
\end{equation}
and the positivity constraints on the fermion spectral weight is
$\mbox{Im} ( G_F ( \omega + i 0^{+} )) > 0 $.
The presence of particle-hole symmetry requires that $\mbox{Im} (G_F
( \omega + i0^{+} ))$ is an even function of $\omega$ - this
simplifies the analysis considerably. The low-frequency limit of
$G_{F}$ can be determined completely:
\begin{displaymath}
G_F ( z ) = \left(\frac{\pi}{4 J^2 n_b}\right)^{1/4} \frac{(-1 +
i)}{\sqrt{z}} + \cdots~~~\mbox{Im}(z) > 0
\end{displaymath}
The dynamical susceptibility is found to have the same low-frequency
dependence as in (\ref{marginal}), with the constant $X$ now given by
\begin{equation}
X = \frac{\sqrt{\pi n_b}}{2J}
\end{equation}
As expected, $X$ is a monotonically increasing function of $n_b$.
A complete solution was obtained numerically and the results are
shown in Fig~\ref{fferm}.

The key unresolved issue in this work is of course the range of
validity of the dynamic susceptibility in Eqn~(\ref{marginal}) - this
is important in determining the significance, if any, of our
results for dynamic neutron scattering experiments on random
antiferromagnets~\cite{birg,gabe,kagome}: ({\em a\/})
What are the consequences
of $1/M$ fluctuations in the infinite-range model $Z_{0}$ (Eqns
(\ref{z0},\ref{qab})) ? This question has been answered for a simpler
infinite-range quantum spin-glass~\cite{ysr} where it was found that
$1/M$ corrections {\em did not modify\/} the low-frequency behavior of the
spectral weight.
The structure of the fluctuations about the present mean-field
theory is much more involved, but it is reasonable to expect that
a similar phenomenon will occur here.
({\em b\/}) Is there an upper-critical dimension above which the
properties of $Z_0$ describe the spin-fluid phase or its phase transition to
spin-glass order in
antiferromagnets with finite-range
interactions ?
({\em c\/})~How are these results
modified in ensembles with a nonzero average $J_{ij}$ ?

We thank A.~Georges, C.M.~Varma, and A.P.~Young for helpful discussions.
This research
was supported
by NSF Grant No. DMR 8857228
and the A.P. Sloan Foundation.

\figure{Spectral weights of $G_B$ and $\chi$ for the bosonic representation
(\ref{bosonrep}) at $\theta = \pi/12$. The `spin' is $n_b$ is order $M$: $n_b =
\kappa M$, and for this value of $\theta$, we found $\kappa = 0.051$.
The sum-rule for $\chi$ is $\int_{0}^{\infty}d \omega \mbox{Im}(\chi(\omega))
= \pi \kappa (1 + \kappa )$ \label{fbos}}

\figure{Spectral weights of $G_F$ and $\chi$ for the fermionic representation
(\ref{fermionrep}). Now the spin $n_b$ is of order unity, and upto rescaling,
the
solution has the same form for all $n_b$. The
sum-rule for $\chi$ is $\int_{0}^{\infty}d \omega \mbox{Im}(\chi(\omega))
= \pi n_b /4$ \label{fferm}}

\end{document}